\documentclass[smallcondensed]{svjour3}     
\smartqed  
\usepackage{amsfonts,amsmath,mathptmx,graphicx}

\begin{document}


\title{The linear optical response of the quantum vacuum}
\subtitle{}
\author{M. Hawton \and L. L. S\'{a}nchez-Soto \and G. Leuchs}
\institute{M. Hawton \at
             Department of Physics, Lakehead University, Thunder Bay,
             ON P7B 5E1, Canada \\
             \email{mhawton@lakeheadu.ca}          
           \and
           L. L. S\'{a}nchez-Soto  \at
              Departamento de \'Optica, Facultad de F\'{\i}sica,
              Universidad Complutense, 28040 Madrid, Spain \\
    Max-Planck-Institut f\"ur die Physik des Lichts, 
               Staudtstra\ss e 2, 91058 Erlangen, Germany\\
                  \email{lsanchez@fis.ucm.es}    
              \and
              G. Leuchs  \at
              Max-Planck-Institut f\"ur die Physik des Lichts, 
               Staudtstra\ss e 2, 91058 Erlangen, Germany \\
                  \email{gleuchs@mpg.mpl.de}        
}

\date{Received: \today / Accepted: date}
\maketitle

\begin{abstract}
  We show that the interpretation of
  $\mathbf{D}=\varepsilon_{0} \mathbf{E}$ as vacuum polarization is
  consistent with quantum electrodynamics. A free electromagnetic field
  polarizes the vacuum but the magnetization and polarization currents
  cancel giving zero source current. The speed of light is a universal
  constant while the fine structure constant that couples the EM field
  to matter runs. In that sense, the quantum vacuum can be understood
  as a modern Lorentz invariant ether.
\end{abstract}

\keywords{Quantum vacuum \and Linear response \and Fine structure constant}


\section{Introduction}

Quantum electrodynamics (QED) is the most successful theory in human
history, but it is normally relegated to short range or high energy
phenomena such as the Lamb shift, collisions of high energy particle
beams and ultra-intense laser fields. Except for photon emission and
absorption and the Casimir effect, its relevance to low-energy
laboratory physics and every day life is unclear~\cite{Milonni,Boi}. 

Recently it has been proposed that
$\mathbf{D=} \varepsilon_{0}\mathbf{E}$ is the vacuum polarization due
to virtual pairs of all types of charged elementary particle in
Nature~\cite{LeuchsSS10,LeuchsSS13,Urban11}.  This is a paradigm shift
in our physical picture of the vacuum, but we will show here that this
interpretation of $ \varepsilon_{0}$ is consistent with QED.

In a dielectric material the electric displacement is
$\mathbf{D} =\varepsilon_{0}\mathbf{E}+\mathbf{P}$, where
$\mathbf{P} =\varepsilon_{0}\chi_{e}\mathbf{E}$ is the polarization
due to the electric field $\mathbf{E}$ and $\chi_{e}$ is the electric
susceptibility. If the material is magnetic the field strength is
$ \mathbf{H}=\mathbf{B}/\mu_{0}-\mathbf{M}$, where $\mathbf{B}$ is the
magnetic flux density and $\mathbf{M}$ is the magnetization. A field
independent dielectric permittivity $\varepsilon $ and magnetic
permeability $\mu $ describe its linear response. The dependence of
$\varepsilon $ and $\mu $ on the frequency plays a central role in
applications and in some situations their dependence on wave vector
can be significant~\cite{Horsley}. The relationship between
$\mathbf{D}$ and $\mathbf{E}$ and between $\mathbf{H}$ and
$\mathbf{B}$ is nonlocal in space and time, but local in reciprocal
spacetime, where a dielectric permittivity
$\varepsilon ( \omega ,\mathbf{k}) $ and magnetic permeability
$\mu ( \omega ,\mathbf{k} ) $ can be defined such that
$\mathbf{D} (\omega, \mathbf{k})  = \varepsilon ( \omega ,\mathbf{k})
\mathbf{E} ( \omega ,\mathbf{k}) $ and $\mathbf{H} ( \omega
,\mathbf{k}) =\mathbf{B}  ( \omega ,\mathbf{k}) /\mu ( \omega ,\mathbf{k}) $.

Consistency with the quantitative predictions of QED
will be maintained here, but $\varepsilon_{0}$\ and $\mu_{0}$\ will
be interpreted as the dielectric permittivity and magnetic permeability of
vacuum. Vacuum has a crucial property that is does not share with dielectric
and magnetic materials; it is Lorentz invariant. Due to this property
Michaelson and Morley failed to detect the motion of the earth through the
``ether'', which in the present context is the quantum vacuum. Empty spacetime
is homogeneous so variations of $\varepsilon_{0}$  and $\mu_{0}$ can occur
only in the presence of charged matter. The linear response of vacuum must
be Lorentz invariant, so in reciprocal space the susceptibility of vacuum
must be a function of $k^{2} = \omega ^{2}/c^{2}-\mathbf{k}^{2}$. 

The condition $k^{2}=0$ describing a freely propagating photon is
referred to as on-shellness in QED. In relativity a particle with mass
$m$ satisfying the dispersion relation
$\omega^{2}=\mathbf{k}^{2}c^{2}+m^{2}c^{4}/\hbar ^{2}$ is referred to
as on-mass-shell or just on-shell. A real photon has zero mass, so for
an on-shell photon $\omega^{2}=\mathbf{k}^{2}c^{2}.$

It is conventional in QED to use natural units with $\varepsilon_{0}$
set equal to $1$ for all values of $k^{2}$. In the dielectric model of
vacuum presented here $\varepsilon_{0}$ and $\mu_{0}$ are functions of
the off-shellness of the photon, $k^{2}$. 

This paper is organized as follows: Section~\ref{sec:vacpol} describes
vacuum polarization due to creation of virtual particle-antiparticle
pairs.  The dielectric model of the quantum vacuum is presented in
Section~\ref{sec:dmol}. The interpretation of $\mathbf{D}$ as vacuum
polarization is discussed in Section~\ref{sec:D}. Finally, our
conclusions are summarized in Section~\ref{sec:con}.

\section{Vacuum polarization}
\label{sec:vacpol}

\begin{figure}[t]
  \centering
  \includegraphics[width=.75 \columnwidth]{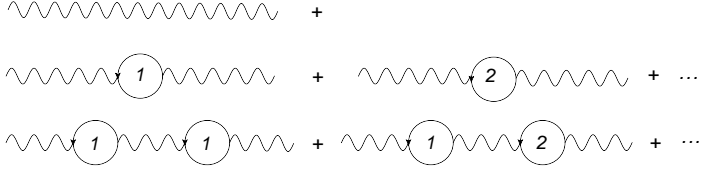}
  \caption{Vacuum polarization in the one-loop approximation. All
    types of virtual pairs in Nature (electron, muon, tau, quarks,
    etc.) polarize the vacuum. This polarization is maximal for a free
    EM field for which $\omega =|\mathbf{k}|c$.}
  \label{feynman}
\end{figure}

The Feynman diagrams in Fig.~\ref{feynman} are a pictorial representation of
vacuum polarization in QED in the one-loop approximation.
The wavy lines represent an electromagnetic (EM) field and a dot is a vertex
where this EM field interacts with the fermions. The loop labelled $1$
represents a virtual electron-positron pair created at space-time
point  $x_{1}= ( ct_{1},\mathbf{x}_{1} ) $ and annihilated at $x_{2}$. The
loop labelled $2$ represents creation of, say, a muon-antimuon pair and so on.
All types of virtual pairs of charged elementary particles in all their
varieties~\cite{LeuchsSS10,LeuchsSS13} are included. 

The time for which a pair can exist and the distance a particle or
antiparticle can travel is limited by the uncertainty principle. When
a virtual pair is created by a photon with frequency $\omega $ its
excess energy is $\Delta E \geq 2mc^{2}-\hbar \omega $ so it can exist
only for a time $t_{2}-t_{1}\leq \hbar /\Delta E$.~\footnote{While a
  real pair cannot be created by absorbing a photon due to
  simultaneous conservation of energy and momentum, this restriction
  does not apply to the ephemeral creation of virtual pairs.}  Similar
restrictions apply to the distance that the virtual particles can
travel.

A loop in Fig.~\ref{feynman} is analogous to an atom in a polarizable
dielectric, with Drude oscillator frequency $2mc^{2}/\hbar$. If its
dipole moment is $d$ and the volume of an atom is $V$, its
polarization is $d/V$. These polarizable virtual atoms fill
space-time. Based on the uncertainty principle and this oscillator
model it was found in Ref.~\cite{LeuchsSS10} that the dielectric
constant of vacuum can be expressed as
\begin{equation}
  \varepsilon_{0}=f\frac{1}{\hbar c} \sum_{j}^{\mathrm{e.\, p.}}q_{j}^{2}  \label{e0}
\end{equation}
where $f$ is a geometrical factor of order unity, $q_{j}$ is charge
and the sum is over all elementary particle types with electric
charge, that is the fermions, the $W$ bosons and whatever surprise
Nature has not yet revealed to us.

In QED the vacuum polarizability is calculated in a standard way in
momentum space as a sum over fermion momenta $p$.  The increase in
density of states with $p$ leads to a geometrical factor $f$ that we
will show is weakly mass dependent. With this modification,
Ref.~\cite{LeuchsSS10} describes a physical model of the QED
process. Vacuum is a polarizable medium very like a material
dielectric but there are some surprises. To maintain Lorentz
invariance $\mathbf{D}$ and $\mathbf{H}$ should form a tensor
proportional to the EM field tensor so $\mathbf{D} =\varepsilon_{0} 
 \mathbf{E}$ and $\mathbf{H}= \varepsilon_{0} c^{2} \mathbf{B}$
implies  that the vacuum must also be magnetizable with
$\mu_{0}( k^{2}) \varepsilon_{0}( k^{2}) =1/c^{2}$. As a consequence
the speed of light is a universal constant and uniform motion of an
observer does not change the constitutive relations or ME. For
on-shell photons the dielectric permittivity $\varepsilon_{0}( 0) $
does not fall-off with frequency as it does in a material
dielectric. This is a consequence the equivalence of all inertial
observers.

The effective fine structure constant that determines the strength of
the photon-matter interaction is
\begin{equation}
  \alpha_{\mathrm{eff}}( k^{2} ) =\frac{e^{2}}{4\pi \varepsilon_{0} (
    k^{2}) \hbar c} \, .  
  \label{alpha}
\end{equation}
This $k^{2}$-dependence of the coupling is called \emph{running}.

\section{The dielectric model of the vacuum}
\label{sec:dmol}

Motivated by the dielectric model of a material, running vacuum
polarization effects will be incorporated into an effective Lagrangian
as in Ref.~\cite{KennedyLynn}.  Renormalization and the calculation of
vacuum polarization in QED can be found in every textbook and is
briefly summarized in Appendix 1. The polarization is a divergent sum
over fermion momenta, so a cut-off $\Lambda$ is required. Modes beyond
this cut-off are incorporated into what is called the bare vacuum. To
facilitate comparison with QED the dielectric permittivity of bare
vacuum is called $\varepsilon_{0} Z_{3}$ and the susceptibility of
vacuum is $\Pi ( k^{2} ) $, where $\Pi ( k^{2} ) $ is the second-order
vacuum polarizability. Here and throughout
$\varepsilon_{0}\equiv \varepsilon_{0}(0) $ and
$\mu_{0}\equiv \mu_{0}( 0)$.

It is proved in Appendix 2 that the fermion-field interaction term can
be written as $\mu_{0} j^{\mu}A_{\mu} = \Pi_{2} ( k^{2} ) (
\varepsilon_{0}\mathbf{E}^{2}-\mathbf{B}^{2}/\mu_{0})$, with
$A^{\mu}= ( \phi /c, \mathbf{A}) $.  Thus, the bare vacuum and
polarization can be combined to give
\begin{equation}
  \frac{1}{2}   [ Z_{3}+\Pi ( k^{2} )  ] 
  ( \varepsilon_{0} \mathbf{E}^{2}-\mathbf{B}^{2}/\mu_{0}) =
  \frac{1}{2} [ \varepsilon_{0} ( k^{2}) \mathbf{E}^{2}
  - \mu_{0}^{-1} ( k^{2} )  \mathbf{B}^{2} ] \, ,
  \label{All}
\end{equation}
where $\mu_{0}^{-1}(k^{2}) /c^{2}=\varepsilon_{0}(k^{2}) $ on the
right-hand side of (\ref{All}) and $c$ is a universal constant as will
be verified at the end of this section. It is then natural to define
the coefficient of
$( \varepsilon_{0} \mathbf{E}^{2}-
\mathbf{B}^{2}/\mu_{0})/\varepsilon_{0} $ as the dielectric
permittivity of vacuum, $\varepsilon_{0}(k^2)$. The physics of vacuum
polarization due to all of the fermions in Nature is in this function,
which for $k^2=0$ is just the well known $\varepsilon_0$ of classical
electromagnetism.\footnote{Standard procedure would be not to change
  the notation when changing the units. In Gaussian units
  $\Pi (k^{2})$ or $Z_{3} + \Pi(k^{2})$ is dimensionless. When going
  to SI units the same expression has the dimension [As/Vm]. Here we
  deviate from this standard procedure, such that $Z_{3} + \Pi(k^{2})$
  has the same value (unity, for $k^{2}=0$) also in the SI units, just
  as in Gaussian units for easier comparison. This, however, requires
  multiplication by the $k=0$ value of $\varepsilon_{0}(k^{2})$, which
  we denote by $\varepsilon_{0}$. We stress that $\varepsilon_{0}$
  does not merely describe conversion from Gaussian to SI units. Just
  like the ``one'' in $Z_{3} + \Pi(k^{2})$ in Gaussian units,
  $\varepsilon_{0}$ has physical significance in that it describes in
  SI units the linear response of the vacuum to an electric field
  under on-shell conditions, $k^{2}=0$, the very response we are
  relating to the properties of the quantum vacuum in this article.}

The separation of $\varepsilon_0(k^2)$ into bare vacuum and
polarization parts is not observable, so it will be rewritten as the
difference between $\varepsilon_0(0)$ and the reduction in vacuum
susceptibility due to off-shellness, $\Delta \Pi(k^2)$; that is,
\begin{equation}
  \varepsilon_{0} ( k^{2} ) \equiv \varepsilon_{0}
  [Z_{3}+\Pi ( k^{2}) ]= \varepsilon_{0} [1-\Delta \Pi ( k^{2}) ].
  \label{epsilon0}
\end{equation}
With this substitution the effective Lagrangian becomes
\begin{equation}
  \mathcal{L}_{\mathrm{eff}} = \frac{1}{2} 
  [ \varepsilon_{0}( k^{2}) \mathbf{E}^{2} -
  \mu_{0}^{-1}(k^{2})  \mathbf{B}^{2} ] - 
  j^{\mu}A_{\mu}+ \mathcal{L}_{\text{\textrm{fer}}}
  \label{EffectiveLagrangian}
\end{equation}
where $j^{\mu}$ is the electric current external to the vacuum and
$\mathcal{L}_{\text{\textrm{fer}}}$ stands for the fermionic part. The
form of (\ref{EffectiveLagrangian}) makes it clear that
$\varepsilon_{0}(k^{2}) $ should be interpreted as the dielectric
permittivity of vacuum.

The dielectric permittivity (\ref{epsilon0}) and the effective
Lagrangian (\ref{EffectiveLagrangian}) are the basis for our
dielectric model. The Lagrange equations derived from
$\mathcal{L}_{\mathrm{eff}}$ are the Dirac equation and, in the Lorenz
gauge, the ME $k^{2}g^{\mu \nu}A_{\nu }=-\mu_{0}( k^{2}) j^{\mu}$.
The latter equation gives the Green function
\begin{equation}
  iD_{F}^{\mu \nu}( k) =-\frac{g^{\mu \nu}}
  {\varepsilon_{0}( k^{2}) ( k^{2}+i\eta )} \, ,
  \label{PhysicalPropagator}
\end{equation}
where $i\eta $ enforces time ordering by choosing the positive sign
for $\eta $ in the contour integral over $\omega $. This is equivalent
to (\ref{DFk}) of Appendix 1, but without the need for summation.

The solution of the Dirac equation in the presence
of the four-potential $A^{\mu} $ gives the
fermion current density
\begin{equation}
j_{\mathrm{fer}}^{\mu} ( k) =-\mu_{0}\Pi ( k^{2})
( k^{2}g^{\mu \nu}-k^{\mu}k^{\nu}) A_{\nu}( k)
\label{Ward}
\end{equation}
due to the Ward identity, which ensures $j_{\mathrm{fer}}^{\mu}$ is a
conserved current. If the calculation is not Lorentz invariant an
on-shell photon is found to have nonzero mass~\cite{Itz,PS} and this is
clearly wrong: Lorentz invariance must be maintained.

To second order in perturbation theory the reduction in
vacuum dielectric permittivity relative to its maximum value at
$k^{2}=0$ is found to be 
\begin{eqnarray}
\varepsilon_{0}\Delta \Pi ( k^{2}) &=&\frac{6}{12\pi ^{2}\hbar c}%
\sum_{j}q_{j}^{2}\int_{0}^{1}dxx( 1-x) \ln \left[ 1-\frac{\hbar
^{2}k^{2}}{m_{j}^{2}c^{2}}x( 1-x) \right]  
\nonumber \\
& \simeq & \frac{1}{12\pi ^{2}\hbar c}\sum_{j}q_{j}^{2}  
 \ln \left ( \frac{\hbar^{2}k^{2}}{A m_{j}^{2}c^{2}} \right ) \, ,
 \label{DeltaPol}
\end{eqnarray}
where the second line is valid when $\hbar^{2}\left\vert
  k^{2}\right\vert \gg m_{j}^{2}c^{2}$ and $A= \exp(5/3)$.
The standard relationship $\mathbf{D}=\varepsilon_{0}\mathbf{E}$ of
classical  electromagnetism is maintained here except that with running
$\varepsilon_{0}( k^{2}) \leq \varepsilon_{0}$.  Since the vacuum is Lorentz
invariant, the linear response of vacuum is described by 
\begin{eqnarray}
\mathbf{D}( k) =\varepsilon_{0}( k^{2}) \mathbf{E}( k) \, ,
\qquad 
\mathbf{H}( k) =c^{2} \varepsilon_{0} (k^{2})  \mathbf{B}( k) \, .  
\label{VacuumH}
\end{eqnarray}
Using (\ref{VacuumH}), Eq. (\ref{Ward}) give the ME 
\begin{eqnarray}
i\mathbf{k\cdot D} = \rho \, , 
  \qquad 
i\mathbf{k} \times \mathbf{H}+i
\frac{\omega}{c}\mathbf{D} =\mathbf{j}.
\label{MEk2}
\end{eqnarray}

The dielectric properties of vacuum differ from those of a material
medium in two important ways: $\ln k^{2}$ dependence replaces the
usual $\omega $ dependence and Lorentz invariance requires that
$\varepsilon_{0}( k^{2}) \mu_{0}( k^{2}) =1/c^{2}$. The speed $c$ is a
universal constant while the impedance
$1/[ c\varepsilon_{0}(k^{2} )] $ that couples the field to matter
runs. On the photon mass shell
$k^{2}=\omega ^{2}/c^{2}-\mathbf{k}^{2}=0$ so a \emph{free photon}
always sees $\varepsilon_{0} $ and there is no running. Since
$ \mathbf{D}=\varepsilon_{0}\mathbf{E}$ and
$\mathbf{H}=\mathbf{B}/\mu_{0}$ both the polarization $\mathbf{D}$ and
the magnetization $\mathbf{H}$ are nonzero. However
$\mathbf{\nabla \times H-\partial D}/\partial t=\mu_{0} \mathbf{j}=0$
as it must for propagation in free space. A free EM wave polarizes and
magnetizes the vacuum but the polarization current exactly cancels the
magnetization current. For the static Coulomb interaction
$ k^{2}=-\mathbf{k}^{2}$ and the interaction strength runs, as
discussed in Appendix~3.

\section{D is vacuum polarization}
\label{sec:D}

A model is described completely by its Lagrangian. In the Lagrangian
(\ref{EffectiveLagrangian}) the dielectric constant of the bare vacuum
is $\varepsilon_{0}Z_{3}=\varepsilon_{0}\left[ 1-\Pi_{2}( 0) \right] .$
This is a purely theoretical construct, since its consequences cannot
be observed. The effective Lagrangian (\ref{EffectiveLagrangian})
describes the physical vacuum with dielectric constant
$\varepsilon _{0}( k^{2}) $ that is accessible to experiment. With
polarization included, the electric displacement is partitioned into
its bare EM part and its vacuum polarization part as
\begin{equation}
  \mathbf{D}( k) =\varepsilon_{0} Z_{3}\mathbf{E}( k)
  +\varepsilon_{0} \Pi ( k^{2} ) \mathbf{E}( k) .
\label{D}
\end{equation}
The parameter $Z_{3}$ depends on the fermion momentum cut-off used to
calculate $\Pi ( k^{2}) $ but $\mathbf{D}( k) $ is
independent of this cut-off.

If the fermion momentum cut-off is the Landau
pole~\cite{Landau,Weinberg}; that is,  if
$\Lambda=\Lambda_{L}$, $Z_{3}=0$ and the dielectric model has some
remarkable properties: the bare vacuum contains no zero-point EM field
and so is truly empty. The electric displacement
$\mathbf{D}=\varepsilon_{0}c\Pi_{2}(k^{2}) \mathbf{E}$ must of course
be exactly the vacuum polarization.  The classical ME in a medium with
charge and current sources are
\begin{align}
& \nabla \cdot \mathbf{B} = 0,  \qquad 
\nabla \times \mathbf{E} 
+\partial \mathbf{B}/\partial t =  0,  \label{Mgeo} \\
& \nabla \cdot \mathbf{D} =\rho ,  \qquad
\nabla \times \mathbf{H}-\partial \mathbf{D}/\partial t =\mathbf{j.}
\label{M4}
\end{align}%
Equations (\ref{Mgeo}) allow us to define the four-vector potential
that drives the creation of the fermion pairs.  If $\Pi =1$,
Eq.~(\ref{M4}) simply says that charge density is the divergence of
polarization and the fermion current is the sum of its polarization
and magnetization parts. In the dielectric model (\ref{M4}) are a
consequence of the Dirac equation. Without vacuum polarization and
magnetization there would be no propagating EM waves.

For a cut-off at the Landau pole all of $\mathbf{D}$ is vacuum
polarization due to all the elementary fermions in Nature. Agreement
with the classical ME requires $\varepsilon_{0}( 0)
=\varepsilon_{0}\Pi ( 0) $,  so the on-shell vacuum
susceptibility is $\Pi ( 0) =1.$ According to (\ref{alpha}) and
(\ref{epsilon0}), at the Landau pole
$\Delta \Pi ( \Lambda_{L}^{2}) =1,$
$\varepsilon_{0}( \Lambda _{L}^{2}) =\varepsilon_{0}\Pi (
\Lambda_{L}^{2}) =0$ and
$ \mathbf{E}=\mathbf{D}/\varepsilon_{0}( k^{2}) $ diverges. To second
order in perturbation theory $\Pi ( 0) =\Pi_{2}( 0) $ is given by
(\ref{Pobservable}) in Appendix 1. The sum over masses can be
evaluated, but for clarity the average of the logarithms of the
fermion rest energies in the Standard Model $mc^{2} \simeq 0.25$~GeV
will be used. Setting $\Lambda =\Lambda_{L}$, Eq.~(\ref{Pobservable})
can be approximated as
\begin{equation}
\varepsilon_{0}\Pi_{2}( 0) =\frac{e^{2}}{12\pi ^{2}\hbar c} 
\ln \left ( \frac{\hbar ^{2}\Lambda_{L}^{2}}{\bar{m}^{2}c^{2}} \right )
 \sum_{j}^{\mathrm{e. \, p.}} \frac{q_{j}^{2}}{e^{2}}.  \label{Pi}
\end{equation}
In the QED-based dielectric model, $\Pi_{2}( 0) $ evaluated using a
cut-off at the Landau pole is exactly $1$, because this is where the
singularity occurs. For the Standard model
$\sum_{j}q_{j}^{2} = 9e^{2}$ so $ \Pi_{2}( 0) =1$ requires
$\hbar c\Lambda_{L}\simeq 10^{30}$~GeV, which on a log scale is close
to $\hbar c\Lambda_{L} \simeq 10^{34}$~GeV calculated for the Standard
Model using QED~\cite{Stuben}. With supersymmetry,
$\hbar c\Lambda_{L}\simeq 10^{20}$ to $10^{17}$~GeV, so the Landau
pole provides a natural cut-off on the Planck scale, where gravity
becomes important. In this case the number of particles is doubled and
the $\ln $ factor is roughly halved. For comparison with the
semiclassical model~\cite{LeuchsSS10,LeuchsSS13}, Eq.~(\ref{e0}) can
be written as $f4\pi \alpha \sum q_{j}^{2}/e^{2}=1.$ The QED based
calculation gives
\begin{equation}
  f= \frac{1}{12\pi^{2}}  
  \ln \left ( \frac{\hbar^{2}\Lambda_{L}^{2}}{\bar{m}^{2}c^{2}} \right ) \, . 
\end{equation}

Zeldovich considered a Lagrangian in which the no electromagnetic term
comes from the interaction of the particle with the field and came to
the same conclusions~\cite{Zeldovich}: in the absence of vacuum
polarization electric and magnetic fields act on Dirac fermions but
there is no field energy and no electromagnetic wave propagation. In
an absolute void it makes no sense to talk about ME
or light propagation.  Only after vacuum polarization is introduced
does the effective Lagrangian give the ME, EM waves travelling at the
speed of light and photons. This theory does not call for the quantum
propagator.

\section{Conclusion}
\label{sec:con}

According to the dielectric model derived here the vacuum is a polarizable
medium with dielectric constant $\varepsilon_{0}( k^{2})
=\varepsilon_{0}[1-\Delta \Pi ( k^{2}) ]\leq \varepsilon_{0}$
where $\Delta \Pi $ is the QED reduction  in vacuum polarizability relative
to the on-shell condition $k^{2}=0.$ Since Lorentz invariance requires $\mu
_{0}( k^{2}) \varepsilon_{0}( k^{2}) =1/c^{2}$ the
speed of light is a universal constant. For a free photon $k^{2}=0$, so there
is no change in $\varepsilon_{0}( k^{2}) $ with photon energy;
it does not run. Since $E=c B \neq 0$, for any free EM wave it will polarize
and magnetize the vacuum, but their contributions to the electric
four-current cancel giving $j^{\mu}=0$.

For a fermion momentum cut-off at the Landau pole $\Lambda_{L}$ the
dielectric model has some remarkable properties. In the absence of vacuum
polarization there is no EM field term, no field energy, no EM wave
propagation, and the electric displacement $\mathbf{D}$ is exactly the
vacuum polarization. The QED model is clearly an oversimplification but the
standard model and SUSY have analogous Landau poles~\cite{Stuben}.

The dielectric model predicts no new observable results but it suggest
a paradigm shift in our physical picture of the vacuum. Any EM field
creates virtual pairs of all charged elementary particles types in
Nature. At the scale of classical EM and quantum optics
$\varepsilon_{0}=\varepsilon_{0}( 0) $ and the vacuum is maximally
polarized.

\section*{Appendix 1. Renormalization in QED}

In QED one starts with an invariant Lagrangian density from which the
conjugate momenta, Hamiltonian and equations of motion can be
derived. The parameters in the bare Lagrangian are not observable so
it is renormalized by rescaling the charges, masses and field
strengths to measurable values.  

The EM term is $\frac{1}{2}( \varepsilon_{0}\mathbf{E}_{0}^{2}-
\mathbf{ B}_{0}^{2}/\mu_{0})$. Rescaling of the four-potential according to
$ A_{0}^{\mu}=\sqrt{Z_{3}}A^{\mu}$ gives
$\frac{1}{2}Z_{3}( \varepsilon_{0}\mathbf{E}^{2}-
\mathbf{B}^{2}/\mu_{0})$.  The relationship between the bare charge
$e_{0}$ and the physical charge $e$ is also rescaled so that
$e_{0}=e/\sqrt{Z_{3}}.$ The Lagrangian is then split into observable
parts and divergent parts called counterterms. The QED vacuum
polarizability to second order of perturbation theory in $e$ is called
$\Pi_{2}( k^{2})$. For photons with $m=0$, the on-shell condition is
$\omega ^{2}=\mathbf{k}^{2}c^{2}$ or $k^{2}=0.$ On-shell
renormalization at $k^{2}=0$ gives $Z_{3}=1-\Pi_{2}( 0)$. The
renormalized EM Lagrangian becomes
$\frac{1}{2}( \varepsilon_{0}\mathbf{E}^{2}-\mathbf{B}^{2}/\mu _{0}) $
and the polarizability is redefined to incorporate the
counterterm. With
\begin{equation}
\Delta \Pi ( k^{2}) \equiv \Pi_{2}( 0) -\Pi
_{2}( k^{2})  \label{DeltaPi}
\end{equation}
this redefined polarizability is positive and equal to zero at the scale of
the classical ME and quantum optics. It is called $\Delta \Pi $ here to
emphasize that it is the \emph{reduction} in polarizability relative to its
value at $k^{2}=0$.

For concreteness the equations used here will be taken from Peskin and
Schroeder~\cite{PS}. The rescaled Lagrangian density
can be written as 
\begin{equation}
\mathcal{L}=\frac{1}{2}Z_{3}( \varepsilon_{0}\mathbf{E}^{2}-\mathbf{B}%
^{2}/\mu_{0}) -j^{\mu}A_{\mu}+\mathcal{L}_{\text{\textrm{fer}}}.
\label{Lrescaled}
\end{equation}
It is conventional in QED to omit $e$ in the definition of $j^{\mu}$ but it
is included here so that $j^{\mu}$ is the electric four-current. Details of 
$\mathcal{L}_{\text{\textrm{fer}}}$ are omitted, because renormalization of
the fermion masses will not be discussed here. The $k$-space current density
due to the four-potential $A^{\nu}=( \Phi /c,\mathbf{A}) $ is of
the form (\ref{Ward}). 

The Dirac equation that can be derived from the renormalized
Lagrangian (\ref{EffectiveLagrangian}) is the basis for the
calculation of $\Pi_{2}( k^{2}) $. Relative to its maximum value at
$k^{2}=0$ the vacuum dielectric permittivity is given by
(\ref{DeltaPol}) with
\begin{equation}
  \varepsilon_{0}\Pi_{2}( 0) =\frac{1}{12\pi ^{2}\hbar c}
  \sum_{j}q_{j}^{2}
  \ln \left ( \frac{\hbar ^{2} \Lambda^{2}}{A m_{j}^{2}c^{2}} \right ) \,  
  \label{Pobservable}
\end{equation}
where $A = \exp (5/3)$. Equation (\ref{DeltaPol}) is a generalization
of (7.91) in \cite{PS} summed to include all fermion types. This sum
is used to provide a simple model but it is recognized that a proper
QED-based calculation should be performed.  Vacuum polarizability
$\widehat{\Pi}_{2}( k^{2}) $ in their (7.91) is called
$- \Delta \Pi ( k^{2}) $ here. In our notation $ \Pi_{2}( 0) >0$ so that
$Z_{3}-1=-\Pi_{2}( 0) $ which is the opposite sign convention to
(10.44) in \cite{PS}. Equation (\ref {DeltaPol}) and
(\ref{Pobservable}) are evaluated to second order in $e$ (the one-loop
approximation), but the exact vacuum polarizability is the sum over
all one-particle irreducible (1PI) Feynman diagrams. (A 1PI diagram is
any diagram that cannot be split in two by removing a single line.)

Equation (\ref{DeltaPol}) is independent of the Lorentz and gauge
independent regularization technique used to integrate over fermion
momenta. Equation~(\ref {Pobservable}) is included to show the
relationship of $\Pi_{2}$ to a cut-off that may be physically
significant. In the Wilson condensed matter analogy~\cite{Wilson}, the
atomic scale provides a natural cutoff. In QED there may also be a
cut-off, possibly due to gravity. The vacuum polarizability $\Pi _{2}$
can be evaluated in a Lorentz and gauge invariant way using
dimensional regularization in $4-\eta $ dimensions. As in \cite{PS}
(12.34), we can define $\ln ( \Lambda^{2}/\mu ^{2}) =2/\eta $
where $ \mu $ is the renormalization scale. In the limit
$\mu \rightarrow 0$ so that $\eta \rightarrow 0$ and the integral is
four-dimensional even for finite $ \Lambda$.

In Fig.~\ref{feynman} a wavy line represents propagation of a bare
photon between points in space-time and the complete diagram describes
the physical photon propagator that takes into account vacuum
polarization. The relativistic propagator is defined as the vacuum
expectation value of field operators at $x_{1}$ and $x_{2}.$ Since the
vacuum is homogeneous it depends only on $x=x_{2}-x_{1}$. The time
ordered Feynman photon propagator is
$ D_{F}^{\mu \nu}( x,0) \equiv \left\langle 0\left\vert \mathcal{T}%
    A^{\mu}( x) A^{\nu}( 0) \right\vert 0\right\rangle .$ Due to its
$A^{\mu}( x) $ dependence it satisfies the homogeneous Maxwell wave
equation except at the time ordering discontinuity where there is a
$\delta ^{4}( x) $ source term. Since $%
D_{F}^{\mu \nu}( x,0) $ is the response to a $\delta $-function source
it is the Green function for ME. In $k$-space in the Lorenz gauge, the
Maxwell wave equation $k^{2}A^{\mu}=-\mu_{0}j^{\mu}$ can be written as
$ k^{2}g^{\mu \nu}A_{\nu}=-\mu_{0}j^{\mu}.$ Since
$\int d^{4}x\delta ^{4}( x) \exp ( ikx) =1$ a source that is localized
in space-time is uniform in $k$-space so the Maxwell Green function
can be written as
\begin{equation}
iD_{F}^{0\mu \nu}( k) =-\frac{g^{\mu \nu}}{k^{2}+i\eta}.
\label{DF}
\end{equation}
When gauge invariance is taken into account the mathematical details are
considerably more complicated and will not be discussed here.

Evaluation of the physical photon propagator according to Fig.~\ref{feynman}
requires summation over all numbers and types of virtual pairs that is 
\begin{equation}
D_{F}^{\mu \nu}( k) =-i\frac{g^{\mu \nu}}{k^{2}}( 1+\Delta
\Pi +\Delta \Pi ^{2}+..) =\frac{-ig^{\mu \nu}}{k^{2}\left[ 1-\Delta
\Pi ( k^{2}) \right]}.  \label{DFk}
\end{equation}%
The full effect of using this physical photon propagator is to replace the
fine structure constant with (7.77) of~\cite{PS}, 
\begin{equation*}
\alpha_{\mathrm{eff}}( k^{2}) =\frac{\alpha}{1-\Delta \Pi
( k^{2})}
\end{equation*}
which is equivalent to (\ref{alpha}). This $k^{2}$-dependence of the
coupling is called \emph{running}. The effective fine structure
constant $ \alpha_{\mathrm{eff}}( k^{2}) $ diverges at the Landau pole
$\Lambda_{L}$, where $\Delta \Pi ( k^{2}) =1$.

Kennedy and Lynn~\cite{KennedyLynn} describe running coupling by
adding the one-loop interaction energies to the bare Lagrangian to
give the effective Lagrangian
\begin{equation}
  \mathcal{L}_{\mathrm{eff}} \simeq 
  \mathcal{L}+\mathcal{L}_{\text{\textrm{one-loop}}} \, .
\end{equation}
The correct diagrammatic expansion of $\mathcal{L}_{\mathrm{eff}}$ is
through 1PI diagrams. $\mathcal{L}_{\mathrm{eff}}$ is written in terms
of the bare charge and field but substitution of
$A_{0}^{\mu}=\sqrt{Z_{3}}A^{\mu}$ and $e_{0}=e/\sqrt{Z_{3}}$ is a
simple rescaling that does not change the EM energy density or the
charged particle interaction energy. KL define the square of the
effective charge $e_{\ast}^{2}( k^{2}) $ for Coulomb interactions and
replace $e_{0}^{2}$ with $e_{\ast}^{2}$ at an experimental
point. Binger and Brodsky show that the effective charge formalism
eliminates inconsistencies due to step functions in calculations of
the grand unification scale~\cite{Binger}. The Lagrangian introduced
here is essentially a rescaled  Lagrangian in which
$ \varepsilon_{0}( k^{2}) $ runs but $e$ is fixed.

\section{Appendix 2. Linear response to the EM force tensor}

This Appendix includes (1) the interpretation of $j_{\text{\textrm{fer}}}$
as the linear response to the EM field tensor, (2) the transformation from $%
4 $-vector to the $3$-vector ME, and (3) the equivalence of the form of the
EM and interaction terms.

(1) The fermion current can be written as the response to the EM field
tensor as follows: If $A^{\mu}( x) =A_{0}^{\mu}e^{-ik_{\nu}x^{\nu}}$,
the EM force tensor is  $F^{\nu \mu}=-ik^{\nu}A^{\mu}+ik^{\mu}A^{\nu}$ and
\begin{equation}
\mu_{0}j_{\text{\textrm{fer}}}^{\mu}( k) = 
i^{2}\Pi (k^{2}) ( k_{\alpha}k^{\alpha}g^{\mu \nu}A_{\nu}-
k^{\mu}k^{\nu}A_{\nu})  = 
 ik_{\nu}\Pi ( k^{2}) ( ik^{\nu}A^{\mu}-
ik^{\mu}A^{\nu})  = -ik_{\nu}\Pi ( k^{2}) F^{\nu \mu}.
\end{equation}
In this form gauge invariance is obvious but note that both terms were
required to give this result~\cite{Plimak}.

(2) The factor $-ik^{\nu}F_{\nu \mu}$ can be written as 
\begin{equation}
-i \left ( 
\begin{array}{cccc}
0 & E_{x}/c & E_{y}/c & E_{z}/c \\ 
-E_{x}/c & 0 & -B_{z} & B_{y} \\ 
-E_{y}/c & B_{z} & 0 & -B_{x} \\ 
-E_{z}/c & -B_{y} & B_{x} & 0
\end{array}%
\right ) 
\left ( 
\begin{array}{c}
\omega /c \\ 
k_{x} \\ 
k_{y} \\ 
k_{z}
\end{array}
\right ) =
\left ( 
\begin{array}{c}
-i\mathbf{k\cdot E}/c \\ 
i\omega E_{x}/c^{2}+i( \mathbf{k\times B})_{x} \\ 
i\omega E_{y}/c^{2}+i( \mathbf{k\times B})_{y} \\ 
i\omega E_{z}/c^{2}+i( \mathbf{k\times B})_{z}
\end{array}
\right )  , \label{A2}
\end{equation}
or
\begin{equation}
-i\mathbf{k\cdot E}=\rho ,\ i \omega \mathbf{E}- 
i\mathbf{k}\times \mathbf{B} =-\mu_{0}\mathbf{j} .  \label{A2MEk}
\end{equation}

(3) The vacuum polarization contribution to the interaction term
$j^{\mu}A_{\mu}$ and the EM term
$-\frac{1}{2\mu_{0}}F_{\nu \mu}F^{\nu \mu}$ are of the same form since
\begin{eqnarray}
-\frac{1}{2\mu_{0}}F_{\nu \mu}F^{\nu \mu} &=&
 \varepsilon_{0}\mathbf{E}^{2}-\mathbf{B}^{2}/\mu_{0}  \nonumber \\
\mu_{0}j_{\text{\textrm{fer}}}^{\mu} &=&
ik_{\nu}\Pi ( k^{2}) F^{\nu \mu}  \nonumber \\
\mu_{0}j_{\text{\textrm{fer}}}^{\mu}A_{\mu}+ 
\mu_{0}j_{\text{\textrm{fer}}}^{\nu}A_{\nu} &=& 
( ik_{\nu}A_{\mu}-ik_{\mu}A_{\nu}) \Pi( k^{2}) F^{\nu \mu} \nonumber \\
\mu_{0}j_{\text{\textrm{fer}}}^{\mu}A_{\mu} &=&-\frac{1}{2}
\Pi (k^{2}) F_{\nu \mu}F^{\nu \mu}   = \Pi ( k^{2}) 
( \varepsilon_{0}\mathbf{E}^{2}-\mathbf{B}^{2}/\mu_{0}) . 
\end{eqnarray}
Thus the interaction term and the EM term in the Lagrangian can be combined
to give (\ref{All}). A factor $\frac{1}{2}$ arises because $j^{\mu}$ is the
four-current induced by the field $A_{\mu}$.

\section{Appendix 3: The Gottfried-Weisskopf dielectric model}

To the best of our knowledge a dielectric model for the vacuum has not
been developed quantitatively in the literature. Gottfried and
Weisskopf~\cite{GW} consider a dielectric model but they define
$\varepsilon_{0}( a) =\varepsilon_{0}$ for some small charge
separation $a\ll \hbar /( m_{e}c) .$ Then $\varepsilon_{0}$ is large
for $r\gg \hbar /( m_{e}c) $ and can diverge so they do not pursue
this. It in the spirit of renormalization in QED to instead define the
relationship between $ \mathbf{D}$ and $\mathbf{E}$ at charge
separations $r\gg \hbar /( m_{e}c) $ where the coupling strength can
be measured. Then $ \varepsilon_{0}( r\gg \hbar /( m_{e}c) ) \equiv 
\varepsilon_{0}$ at large distances or, equivalently, small momenta, 
where the vacuum is maximally polarized.

This is effectively what is done in conventional QED since
$\frac{1}{2}\varepsilon_{0}\mathbf{E}_{0}^{2} $ is written as
$\frac{1}{2}Z_{3}\varepsilon_{0}\mathbf{E}^{2} $so that
$\varepsilon_{0}$ is replaced with $Z_{3}\varepsilon_{0}$ at the bare
scale. The field is rescaled but the energy density is not changed by
this rescaling. Since
$\alpha =e_{0}^{2}/( 4\pi \varepsilon _{0}\hbar c) =e^{2}/( 4\pi
Z_{3}\varepsilon_{0}\hbar c) \ $ this rescaling implies
$e^{2}=e_{0}^{2}Z_{3}.$ With polarization included the dielectric
permittivity is $\varepsilon_{0}$ at the physical scale. The
dielectric model is consistent with, but does not require,
propagators.

\begin{figure}[tbp]
\centering
\includegraphics[height=5cm]{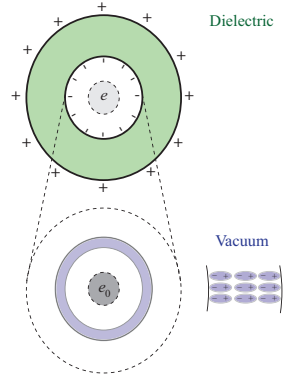} 
\caption{Screening in vacuum due to virtual particle-antiparticle
  pairs.}
\label{GW}
\end{figure}

OKThe vacuum behaves like a material dielectric except that
$\varepsilon_{0}( k^{2}) =\varepsilon_{0}$, where the vacuum is
maximally polarized so that $\Delta \Pi $ is the \emph{reduction} in
polarizability, as sketched in Fig.~\ref{GW}. The green area
represents a dielectric spherical shell such that the charge inside a
sphere of radius $r$ is $q_{\mathrm{in}}=e/\varepsilon_{0} $ due to
the negative surface charge on the inner surface of the green
shell. The expanded region is $r<\hbar /m_{e}c$. By Gauss's theorem
$\Phi =q_{\mathrm{in}}/4\pi \varepsilon_{0}(r) r$ where $q_{in}$ is
the charge inside a sphere of radius $r$.  Because $q_{\mathrm{in}}$
is $r$-dependent there is a net charge inside a spherical shell of
thickness $dr$ equal to $q_{in}( r+dr) -q_{in}( r) $, so the charge
density is nonzero for $r<m_{e}^{-1}.$ Since
$ \varepsilon_{0}^{-1}( k^{2}) $ is not a constant, in $r$-space
$ \varepsilon_{0}^{-1}$ is nonlocal. It is not, strictly speaking,
correct to write the potential as $\Phi =e/4\pi \varepsilon_{0}( r) r$
as in Ref.~\cite{GW}. Quantitative calculations are more easily
performed in $k$-space.

Running of $e^{2}/\varepsilon_{0}( k^{2}) $ is in most ways
equivalent to running of the square of effective charge in conventional QED,z
but the physical interpretation is different. In a dielectric it is possible
to have $\varepsilon_{0}<0,$ but $e_{\mathrm{eff}}^{2}<0$ makes no physical
sense.

The simplest example of running coupling is a static charge, say $+e.$ In
the Coulomb gauge $\mathbf{E}= -\nabla \Phi $ and $\omega
=0$. Equation (\ref{MEk2})  with $k^{2}=-\mathbf{k}^{2}$ then gives 
\begin{equation}
\Phi ( \mathbf{k}^{2} )= \frac{e}
{\mathbf{k}^{2} \varepsilon_{0}( \mathbf{k}^{2})} \, .  
\label{CoulombDielectric}
\end{equation}
In $\mathbf{r}$-space 
\begin{equation}
\Phi ( r) =\int \frac{d^{3}\mathbf{k}}{( 2\pi )^{3}} 
\frac{e}{\mathbf{k}^{2}\varepsilon_{0}( \mathbf{k}^{2})}
\exp( i\mathbf{k}\cdot \mathbf{r}) \, .  
\label{Coulomb_r}
\end{equation}%
This is equivalent to (7.93) in \cite{PS} that is based on (\ref{alpha}).

For electron-positron pairs alone $\Delta \Pi \ll 1$ so that 
$\varepsilon_{0}^{-1}\simeq (1+\Delta \Pi )/\varepsilon_{0}$
and~\cite{LifshitzPitaevskii} 
\begin{equation}
\Phi  ( r ) \simeq \frac{e}{4\pi \varepsilon _{0}r} \times 
 \left \{  
\begin{array}{ll} 
1+\frac{2\alpha }{3\pi } 
\ln \left( \frac{\hbar }{m_{e}cr}\right) -\gamma - \frac{5}{6} 
  & \qquad r \ll \hbar /m_{e}c, \\
& \\
1+\frac{\alpha }{4\sqrt{\pi}} \frac{e^{-2m_{e}cr}/\hbar}
{\left( m_{e}cr/\hbar \right)^{3/2}} 
& \qquad r \gg \hbar /m_{e}c,
\end{array}
\right .
 \label{approxCoulomb}
\end{equation}
where $m_{e}$ is the mass of the electron, $r$ is the distance from the
fixed charge and $\gamma =0.577$ is Euler's constant. The dielectric
constant $\varepsilon_{0}$ decreases with increasing $\mathbf{k}^{2}$ or
decreasing $r.$ For $r<\hbar /m_{e}c$ the Coulomb interaction becomes
stronger as the charges approach each other.



\end{document}